\documentclass[11pt]{article}
\usepackage{amssymb,amsmath,graphicx,citesort,ils08eng}
\begin{document}

\begin{center}

{\LARGE\textbf{Higgs boson decay into bottom quarks and
uncertainties of perturbative QCD predictions}}
\vskip7.5mm
{\large A.L.~Kataev}
\vskip5mm
\textit{Institute for Nuclear Research of Russian Academy of
Sciences
  \\ 117312, Moscow, Russia}
\vskip7.5mm
{\large V.T.~Kim }
\vskip5mm
\textit{St.-Petersburg Nuclear Physics Institute of Russian Academy
of Sciences
  \\ 188300, Gatchina, Russia}
\vskip10mm
\textbf{Abstract}
\vskip2.5mm
\parbox[t]{110mm}{{\small Different methods for treating  the
results of higher-order perturbative QCD calculations of the decay
width of the Standard Model Higgs boson  into bottom quarks are
discussed.
Special attention is paid to  the analysis   of the $\text{M}_\text{H}$
dependence of the decay width $\Gamma(\rm{H}\rightarrow
\rm{\bar{b}b})$ in the cases when the mass of $\rm{b}$-quark is
defined as the running parameter in the $\rm{\overline{\text{MS}}}$-scheme
and as the quark pole mass. The relation between running and pole
masses is taken into account in the order
$\alpha_s^4$-approximation. Some special features of applications of
Analytical Perturbation Theory (APT) are commented.}}
\end{center} \vskip5mm \section{Introduction} \label{Sect} The study
of the Higgs boson decay width into bottom quarks is rather
important for calculations of the branching ratios of this important
ingredient of the Standard Model and its various  extensions.
Current LEP and Tevatron fits of the Standard Model parameters yield
the value for the Higgs boson mass around 
$\text{M}_\text{H} = 76 {+33 \atop -24}$ GeV C.L. 68$\%$. 
With the direct LEP search limit 
$\text{M}_\text{H} \geq 115$~GeV 
the fits provide at C.L. 95\%
$\text{M}_\text{H} \leq 182$~GeV. 
In the case, if the scalar Higgs particle has the mass 
in the region 115 $\text{GeV}\leq \text{M}_\text{H} \leq 2\,{\rm M_W}$, 
its decay width to bottom quarks 
$\Gamma(\rm{H\rightarrow\bar{b}b})$
dominates over other channels. 
In particular, it is determining the branching ratio 
of $\rm{H\rightarrow\gamma\gamma}$ process
which is considered to be one of the most promising channel 
for searches of Higgs particles at LHC 
in the mass region specified above 
(for reviews see~\cite{Spira:1997dg}--\cite{Djouadi:2005gi}).
There is also a possibility that the signal for 
$\rm{H\rightarrow\bar{b}b}$ process may be seen at Tevatron
~\cite{SoldnerRembold:2008ak} through WH- and ZH- channels and at
CMS-TOTEM \cite{Albrow:2006xt} or/and FP420 \cite{Albrow:2005ig}
experimental proposals at LHC, aimed at searches of central
exclusive $\rm{H}$ production, as discussed from theoretical point
of view, e.g., in Refs.~\cite{DeRoeck:2002hk,Petrov:2003yt}.
The mentioned experimentally-oriented motivation is pushing ahead
the intention to study in more detail the dominant theoretical
effects to $\Gamma(\rm{H\rightarrow\bar{b}b})$ in the
region of relatively light Higgs boson. These effects are related to
high-order perturbative QCD predictions with their intrinsic
uncertainties. Moreover, the comparison of various representations
for $\Gamma(\rm{H\rightarrow\bar{b}b})
 =\Gamma_{\rm H\bar{b}b}$ is rather important for planning the experimental
program of high energy linear $\rm{e^+e^-}$-colliders for measuring
Higgs boson couplings~\cite{Djouadi:2005gi}. 

\section{QCD expressions for $\Gamma_{\rm H\bar{b}b}$} 
 \label{Sect1} 
 Let us first consider QCD theoretical prediction  for  
$\Gamma_{\rm H\bar{b}b}$ 
expressed in terms of running $\rm{b}$-quark mass
and the QCD coupling constant in the $\rm{\overline{\text{MS}}}$-scheme as
\begin{equation} \label{MS}
 \Gamma_{\rm{H\bar{b}b}}
  =\Gamma_0^{\text{b}}\,
    \frac{\rm\overline{m}_b^2(\text{M}_\text{H})}
         {\rm m_b^2}\,
     \bigg[1+\sum_{i\geq 1}
              \Delta{\rm \Gamma_i}\,
               a_s^i(\text{M}_\text{H})
     \bigg]\,.
\end{equation}
Here
$\Gamma_0^{\text{b}}=3\sqrt{2}/{8\pi}\rm{G_F}\text{M}_\text{H}\,\text{m}_\text{b}^2$,
$\rm{m_b}$ and $\text{M}_\text{H}$ are the pole $\rm{b}$-quark  and Higgs
boson masses,  $a_s(\text{M}_\text{H})=\alpha_s(\text{M}_\text{H})/\pi$ and $\rm
\overline{m}_b(\text{M}_\text{H})$ are the QCD  running parameters, defined
in the $\rm\overline{\text{MS}}$-scheme. The coefficients
$\Delta\rm{\Gamma}_i$ are known analytically up to 4-th order
correction of perturbation theory~\cite{Baikov:2005rw}. They consist
of  the positive contributions $\rm{d_i^{E}}$, calculated directly
in the Euclidean region, and from the proportional to $\pi^2$
kinematic effects, which appear as the result of analytical
continuation from the Euclidean space-like  to the Minkowskian
time-like region. This $\pi^2$-term arises first in Eq.\ (\ref{MS}) 
at the $a_s^2$-correction~\cite{Gorishnii:1983cu}. Its coefficient
was corrected later  in~\cite{Gorishnii:1990zu},
~\cite{Gorishnii:zr}, but the kinematic $\pi^2$-term remained
unaffected. 
Using the notations  of Ref.~\cite{Chetyrkin:1997wm}
one can write down the following relations:
\begin{eqnarray}
 \Delta{\rm\Gamma_1}
  \!&\!=\!&\!
     \text{d}_1^\text{E}=\frac{17}{3}\,;
 \label{1}\\
 \Delta{\rm\Gamma_2}
  \!&\!=\!&\!
     \text{d}_2^\text{E}-\gamma_0(\beta_0+2\gamma_0)\pi^2/3\,;
 \label{2}\\
 \Delta{\rm\Gamma_3} 
  \!&\!=\!&\!
     \text{d}_3^\text{E}-\big[d_1^{E}(\beta_0+\gamma_0)(\beta_0+2\gamma_0)
    +\beta_1\gamma_0+2\gamma_1(\beta_0+2\gamma_0)\big]\pi^2/3\,;
 \label{3}\\
 \Delta{\rm\Gamma_4}
  \!&\!=\!&\!
     \text{d}_4^{\rm E}-\big[\text{d}_2^{E}(\beta_0+\gamma_0)(3\beta_0+2\gamma_0)
    +\text{d}_1^{E}\beta_1(5\beta_0+6\gamma_0)/2 
    \nonumber \\
  \!&\!+\!&\!{}
     4\text{d}_1^{E}\gamma_1(\beta_0+\gamma_0)
    +\beta_2\gamma_0+2\gamma_1(\beta_1+\gamma_1)
    +\gamma_2(3\beta_0+4\gamma_0)\big]\pi^2/3
    \nonumber \\
  \!&\!+\!&\!{}
     \gamma_0(\beta_0+\gamma_0)(\beta_0+2\gamma_0)
             (3\beta_0+2\gamma_0)\pi^4/30\,,~ 
  \label{4}
\end{eqnarray} 
where the ${\rm n_f}$-dependence of $d_i^\text{E}$
($i>2$) read 
\begin{eqnarray} \nonumber
 \text{d}_2^\text{E}
  \!&\!=\!&\! 
   \left[\frac{10801}{144}
        -\frac{39}{2}\,\zeta_3 \right] 
  -{\rm n_f}
    \left[\frac{65}{24}
         -\frac{2}{3}\,\zeta_3
    \right]\\
  \!&\!\approx\!&\!
    51.567
   -1.907\,{\rm n_f}
  \approx 42.032~~~({\rm n_f=5})\,;
\label{2a} 
\end{eqnarray} 
\begin{eqnarray}
 \nonumber
  \text{d}_3^\text{E}
   \!&\!=\!&\!
    \left[\frac{163613}{5184}
         -\frac{109735}{216}\,\zeta_3
         +\frac{815}{12}\,\zeta_5
    \right]
   -{\rm n_f}
    \left[\frac{46147}{486}
         -\frac{262}{9}\,\zeta_3
         +\frac{5}{6}\,\zeta_4
         +\frac{25}{9}\,\zeta_5
    \right] \\ \nonumber 
  \!&\!{+}\!&\!
   {\rm n_f^2}
    \left[\frac{15511}{11664}
         -\frac{1}{3}\,\zeta_3
    \right] \\
  \!&\!\approx\!&\!
     648.71
    -63.742\,{\rm n_f}
    +0.92913\,{\rm n_f^2}
  \approx 353.23~~~({\rm n_f=5})\,;
 \label{3a}
\end{eqnarray} 
\begin{eqnarray}
 \nonumber
  \text{d}_4^\text{E}
   \!&\! =\!&\!
    \bigg[\frac{10811054729}{497664}
         -\frac{3887351}{324}\,\zeta_3
         +\frac{458425}{432}\,\zeta_3^2
         +\frac{265}{18}\,\zeta_4
         +\frac{373975}{432}\,\zeta_5 \\ \nonumber
  \!&\!\!&\!
         -\frac{1375}{32}\,\zeta_6
         -\frac{178045}{768}\,\zeta_7
   \bigg ] \\ \nonumber
  \!&\!{+}\!&\!  
   {\rm n_f}
    \bigg[-\frac{1045811915}{373248}
          +\frac{5747185}{5184}\,\zeta_3
          -\frac{955}{16}\,\zeta_3^2
          -\frac{9131}{576}\,\zeta_4
          +\frac{41215}{432}\,\zeta_5 \\ \nonumber
  \!&\!\!&\!
          +\frac{2875}{288}\,\zeta_6
          +\frac{665}{72}\,\zeta_7
    \bigg]\\ \nonumber
  \!&\!{+}\!&\! 
   {\rm  n_f^2}
    \left[\frac{220313525}{2239488}
         -\frac{11875}{432}\,\zeta_3
         +\frac{5}{6}\,\zeta_3^2
         +\frac{25}{96}\,\zeta_4
         -\frac{5015}{432}\,\zeta_5
    \right] \\ \nonumber
  \!&\!{+}\!&\!
   {\rm  n_f^3}
    \left[-\frac{520771}{559872}
          +\frac{65}{432}\,\zeta_{3}
          +\frac{1}{144}\,\zeta_{4}
          +\frac{5}{18}\,\zeta_{5}
    \right] \\
  \!&\!\approx\!&\!
    9470.8
   -1454.3\,{\rm n_f}
   +54.783\,{\rm  n_f^2}
   -0.45374\,{\rm n_f^3}
    \approx 3512.2~~~({\rm n_f=5})\,.
 \label{4a} 
\end{eqnarray}
 The term of Eq.\ (\ref{1}) was evaluated   in
Ref.~\cite{Gorishnii:1983cu}. It is in agreement with the expressed
in other ways results of previous studies, performed in
\cite{Becchi:1980vz,Sakai:1980fa,Inami:1980qp}. The
second coefficient was corrected  in~\cite{Gorishnii:1990zu,Gorishnii:zr}. 
The result of Eq.~(\ref{3a}) was  obtained in
Ref.~\cite{Chetyrkin:1996sr}. The exact value of the Euclidean
coefficient of  Eq.~(\ref{4a}), analytically calculated in
~\cite{Baikov:2005rw}, turned out to be in reasonable  agreement
with the estimates, obtained within the used in
Ref.~\cite{Chetyrkin:1997wm} a variant of the effective charge
approach (ECH) and the principle of minimal sensitivity (PMS)
approach. The variant of the two approaches was developed in
Ref.~\cite{Kataev:1995vh}.
The coefficients $\beta_i$ and $\gamma_i$ enter the expansions of
the QCD renormalization group (RG)  $\beta$-function and anomalous
dimension of mass function $\gamma_m$. The QCD $\beta$-function can
be defined as
\begin{eqnarray}
 \label{beta}
  \frac{da_s}{d\ln\mu^2}
   \!&\!=\!&\!
    \beta(a_s) \\ \nonumber
   \!&\!=\!&\!
    -\beta_0\,a_s^2
    -\beta_1\,a_s^3
    -\beta_2\,a_s^4
    -\beta_3\,a_s^5
    -\beta_4\,a_s^6
    +O(a_s^7)\,.
\end{eqnarray}
Its  $\rm{\overline{\text{MS}}}$-scheme expressions  were calculated
analytically up to   4-loop corrections~\cite{van Ritbergen:1997va},
confirmed recently in the work~\cite{Czakon:2004bu}. 
We present here the results of analytical
evaluation of the coefficients of Eq.\ (\ref{beta}) ($\beta_0$ and
$\beta_1$ are scheme-independent) in the   $\rm{\overline{\text{MS}}}$-scheme,
supplemented by the numerical expressions, related to ${\rm n_f}=5$
number of active flavours: 
\begin{eqnarray}
 \nonumber
  \beta_0
   \!&\!=\!&\!
    \frac{1}{4}
     \bigg[11-\frac{2}{3}\,{\rm n_f}\bigg]  \\
   \!&\!\approx\!&\! 
    2.75-0.1667~{\rm n_f}~\approx  1.9167~~~(\rm{n_f}=5) \\
 \nonumber
  \beta_1
   \!&\!=\!&\!
    \frac{1}{16}
     \bigg[102-\frac{38}{3}\,{\rm n_f}\bigg]  \\
   \!&\!\approx\!&\!
    6.375-0.7917\,{\rm n_f} \approx 2.4167~~~({\rm n_f}=5) \\
 \nonumber
  \beta_2
   \!&\!=\!&\!
    \frac{1}{64}
     \bigg[\frac{2857}{2}
          -\frac{5033}{18}\,{\rm n_f}
          +\frac{325}{54}\,{\rm n_f^2}
     \bigg] \\
   \!&\!\approx \!&\!
    22.32-4.3689\,{\rm n_f}+0.09404\,{\rm n_f^2} \approx
     2.8267~~~({\rm n_f}=5) \\ \nonumber
  \beta_3
   \!&\!=\!&\!
    \frac{1}{256}
     \bigg[\bigg(\frac{149753}{6}+3564\,\zeta_3\bigg)
          -\bigg(\frac{1078361}{162}+\frac{6508}{27}\,\zeta_3\bigg)\,{\rm n_f}
  \\ \nonumber
   \!&\!\!&\!
          +\bigg(\frac{50065}{162}+\frac{6472}{81}\,\zeta_3\bigg)\,{\rm n_f^2}
          +\frac{1093}{729}\,{\rm n_f^3}
     \bigg] \\
   \!&\!\approx\!&\!
    114.23-27.134\,{\rm n_f}
          +1.5824\,{\rm n_f^2}
          +0.0059{\rm n_f^3}
    \approx 18.852~~~({\rm n_f}=5)\,.
\end{eqnarray} 
Throughout this
work we fix ${\rm n_f}=5$ and neglect the contribution of lighter
four quarks   to the relation between running mass
$\rm{\overline{m}_b(\text{M}_\text{H})}$ and pole (or on-shell) mass
 ${\rm m_b}$ in Eq.\ (1)  (more details
will be given below).  
This is done for self-consistency of further analysis. 
Indeed, 
the similar contributions to the coefficient function 
of Eq.\ (1) are still unknown 
and are expected to be small.
Notice an interesting fact:  the  growth of the coefficients of
$\beta$-function at ${\rm n_f}=5$ is starting to manifest itself
from the four-loop only  (on the contrary to the case with ${\rm
n_f}=3$ when the values of the coefficients of perturbative series
for the $\beta$-function are
 monotonically increasing  from  the one-loop order).
As to the  five-loop coefficient  $\beta_4$ in Eq.\ (\ref{beta}), it
was estimated in Ref.~\cite{Ellis:1997sb} by means    of Pad\'e
approximations approach. The input information, used in these
estimates, is the analytical result for  the  ${\rm
n_f}^4$-contribution to  $\beta_4$ calculated in~\cite{Gracey:1996up}. 
In our normalization conditions it  has the
following form \begin{equation}
\beta_4^{[4]}=\frac{1}{1024}\bigg[\frac{1205}{2916}-
\frac{152}{81}\,\zeta_3\bigg] {\rm n_f^4}=-0.0017969{\rm n_f^4}
 \end{equation}
The  approximation  of Ref.~\cite{Ellis:1997sb}, which is taking it
into account, reads: 
\begin{eqnarray}
 \label{beta4}
  \beta_4
  \!&\!\approx\!&\!
   \frac{10^5}{1024}
    \bigg[7.59
         -2.19\,{\rm n_f}
         +20.5\,{\rm n_f^2}
         -49.8~{10}^{-5}\,{\rm n_f^3}
         -1.84~{10}^{-5}\,{\rm n_f^4}
    \bigg]
\\ \nonumber 
  \!&\!\approx\!&\! 
   741.2-213.87\,{\rm n_f}
        +20.02\,{\rm n_f^2}
        -0.0483\,{\rm n_f^3}
        -0.0018\,{\rm n_f^4}
    \approx 165.2~~({\rm n_f}=5)\,.
\end{eqnarray} 
We will use this estimate in our  studies.
The related to the $\overline{\rm MS}$-scheme anomalous mass
dimension function
 is defined as
\begin{eqnarray}
 \label{mass}
  \frac{d{\ln}\overline{\rm m}_b}
       {d{\ln}\mu^2}
  \!&\!=\!&\!
   \gamma_m(a_s) \\ \nonumber 
  \!&\!=\!&\!
       -\gamma_0\,a_s
       -\gamma_1\,a_s^2
       -\gamma_2\,a_s^3
       -\gamma_3\,a_s^4
       -\gamma_4\,a_s^5
       +O(a_s^6)\,.
\end{eqnarray}
 The four-loop correction was
calculated independently in Ref.~\cite{Chetyrkin:1997dh} and in
Ref.~\cite{Vermaseren:1997fq}. Let us present the explicitly known
coefficients: 
\begin{eqnarray}
\gamma_0
 \!&\!=\!&\! 1\,; \\
\gamma_1
 \!&\!=\!&\!\frac{1}{16}\bigg[\frac{202}{3}
           -\frac{20}{9}\,{\rm n_f}\bigg] 
  \approx 4.2083-0.13889\,{\rm n_f} 
  \approx 3.5139~~({\rm n_f}=5)\,;
\\ \nonumber
\gamma_2
 \!&\!=\!&\!\frac{1}{64}
             \bigg[1249
                  -\bigg(\frac{2216}{27}
                        +\frac{160}{3}\,\zeta_3
                   \bigg)\,{\rm n_f}
                  -\frac{140}{81}\,{\rm n_f^2}
             \bigg] \\
 \!&\!\approx\!&\! 19.516
                  -2.2841\,{\rm n_f}
                  -0.027006\,{\rm n_f^2}
      \approx 7.42~~(\rm{n_f}=5)\,;
\\ \nonumber
\gamma_3
 \!&\!=\!&\!\frac{1}{256}
             \bigg[\frac{4603055}{162}
                  +\frac{135680}{27}\,\zeta_3
                  -8800\,\zeta_5 
\\ \nonumber 
 \!&\!\!&\!       -\bigg(\frac{91723}{27}
                  +\frac{34192}{9}\,\zeta_3
                  -880\,\zeta_4
                  -\frac{18400}{9}\,\zeta_5
                   \bigg)\,{\rm n_f}
\\ 
 \!&\!\!&\!       +\bigg(\frac{5242}{243}
                        +\frac{800}{9}\,\zeta_3
                        -\frac{160}{3}\,\zeta_4
                   \bigg)\,{\rm n_f^2}
                  -\bigg(\frac{332}{243}
                        -\frac{64}{27}\, \zeta_3
                   \bigg)\,{\rm n_f^3}
             \bigg]
\\ \nonumber 
 \!&\!\approx\!&\!98.933
                 -19.108\,{\rm n_f}
                 +0.27616\,{\rm n_f^2}
                 +0.005793\,{\rm n_f^3} 
     \approx 11.034~~({\rm n_f}=5)\,. 
\end{eqnarray}
In the same work \cite{Ellis:1997sb} the
following model for the  five-loop coefficient 
of $\gamma_m$ 
was proposed:
\begin{equation} \label{gamma4} \gamma_4 \approx  530
-143\,{\rm n_f}+6.67\,{\rm n_f^2}+0.037\,{\rm n_f^3}-8.54\times
10^{-5}\,{\rm n_f^4} \approx
-13.68~~~(\rm{n_f}=5)
\end{equation} It is based on application of
 the variant of the Pad\'e approximation
approach, used for getting  Eq.\ (\ref{beta4})  discussed above. The
explicit analytical expression  for the ${\rm n_f^4}$ contribution
to $\gamma_4$, extracted from the QED  results of Ref.
\cite{PalanquesMestre:1983zy} and   confirmed later on in~
\cite{Gracey:1996up}, namely 
\begin{equation}
 \label{gamma44}
  \gamma_4^{[4]}=
   \frac{1}{12}
    \bigg(\frac{65}{5184}
         +\frac{5\,\zeta_3}{324}
         -\frac{\zeta_4}{36}
    \bigg)\,{\rm n_f^4}
\end{equation}
was used. 
Notice that the numerical value of Eq.\ (\ref{gamma4}) is negative. 
This means  that the uncertainties of this Pad\'e   estimate are not
small. Indeed, the analytical  calculations of   ${\rm n_f^3}$
contribution to Eq.\ (\ref{gamma4}), performed in
Ref.~\cite{Ciuchini:1999cv}, gave 
\begin{equation}
 \label{gamma43}
  \gamma_4^{[3]}
   =\frac{1}{12}
     \bigg[\frac{331865}{124416}
          +\frac{803}{432}\,\zeta_3
          +\frac{7}{12}\,\zeta_4
          -\frac{20}{9}\,\zeta_5
     \bigg]\,{\rm n_f^3}
   =0.10832\,{\rm n_f^3}\,.
\end{equation} 
It does  not agree with the similar coefficient 
in  Eq.\ (\ref{gamma4}). 
Substituting Eq.\ (\ref{gamma43}) into   Eq. (\ref{gamma4}) 
one can  see 
that for ${\rm n_f=5}$ 
the estimate  for $\gamma_4$ is still negative, 
but rather small, 
namely  $\gamma_4\approx-4.76595$. 
We will incorporate this value in our further  analysis 
just by fixing some parts of existing five-loop ambiguities. 
We hope  that this expression, 
obtained by matching the explicit results of Eq.\ (\ref{gamma44}) 
and Eq.\ (\ref{gamma43}) 
and the Pad\'e resummation technique, 
may be improved in the future.

\section{Analytical continuation effects  and APT approach}
Consider   now in more detail  the contributions to
$\Gamma_{\rm{H\bar{b}b}}$ from    the Minkowskian coefficients,
defined in   Eqs.\ (3)--(5). Let us remind, that they  are composed
from the Euclidean terms (see   Eqs.\ (6)--(8))  and the
combinations of the coefficients of  the QCD RG functions
$\beta(a_s)$ and $\gamma_m(a_s)$, which are defining proportional to
$\pi^2$ analytical continuation effects  in Eqs.\ (3)--(5). These
kinematic effects turn out to  be negative and quite sizable. Thus,
the coefficients  $\Delta{\rm\Gamma_i}$ ($i\geq 2$) are smaller,
than their Euclidean ``analogs'' $\text{d}_i^\text{E}$. Indeed, for
${\rm n_f}=5$ we have \cite{Gorishnii:1990zu} 
\begin{equation}
 \Delta{\rm\Gamma_2}\approx 29.147\,. 
\end{equation} 
The numerical
values of other terms obey the same pattern~\cite{Baikov:2005rw}:
\begin{eqnarray}
\Delta{\rm\Gamma_3}\!&\!\approx\!&\!41.758\,;  \\
\Delta{\rm \Gamma_4}\!&\!\approx\!&\!-825.75\,. 
\end{eqnarray} 
This means 
that it is of  interest how the effects of analytical continuation
influence other  results of perturbative QCD predictions. 
In the beginning of 80s 
this problem was discussed 
in the number of works on the subject 
(see, e.g.,~\cite{Krasnikov:1982fx,Radyushkin:1982kg}). 
At that  time the problem of resummation of the $\pi^2$-contributions 
to $\Gamma_{\rm{H\bar{b}b}}$ was also considered 
in Ref.~\cite{Gorishnii:1983cu}. 
However, the real interest to resummations 
of the analytical continuation effects 
was attracted later on after appearance of Contour Improved  Technique 
(CIT)~\cite{Pivovarov:1991rh,Le Diberder:1992te} 
and  Analytic Perturbation Theory (APT), 
in particular. 
This method was proposed and developed 
by D.~V.~Shirkov and I.~L.~Solovtsov in the process of common investigations 
in Refs.~\cite{Shirkov:1997wi}--\cite{Shirkov:2006gv}
and in the separate publications as well 
(see works done with the decisive  contribution
 of I.~L.~Solovtsov,~\cite{Milton:1996fc,Nesterenko:2001xa}, 
 and the ones created by inspiration of D.V.\,Shirkov
 \cite{Shirkov:2000qv,Shirkov:2005sg}). 
This QCD approach was already used 
in various applications (see, e.g.,~\cite{Milton:1997us,Cvetic:2008bn}). 
Among them are the studies of Higgs boson decay 
into ${\rm\bar{b}b}$-pair with the help 
of Fractional Analytical Perturbation Theory (FAPT)~\cite{Bakulev:2006ex}, 
which are complementary to definite considerations 
of Ref.~\cite{Broadhurst:2000yc}. 
Quite recently some new~\cite{Bakulev:2008qq}   
and even a bit corrected \cite{Stefanis:2008dz} discussions 
of applications of FAPT  to $\Gamma_{\rm{H\bar{b}b}}$ 
and their comparisons with the results of Ref.~\cite{Broadhurst:2000yc},  
appeared in the literature. 
In view of this,  
we will focus ourselves here  on the brief discussions 
of several ideas of Ref.~\cite{Broadhurst:2000yc}.
It is worth to stress, that this work was motivated in part  by the
desire to understand whether the bridge may be built between the
renormalon approach (for a review see,  e.g., \cite{Beneke:2000kc})
and the resummation of the proportional to $\beta_0$-effects within
Shirkov--Solovtsov APT. Note, that one of the main cornerstones of
renormalon approach   is $\beta_0$-resummation procedure as well.
Today, thanks  to the works of 
Refs.~\cite{Bakulev:2006ex,Stefanis:2008dz} it is understood, 
that this bridge does exist. 
The essential point in the studies of Ref.~\cite{Broadhurst:2000yc} 
is that the  CIT~\cite{Pivovarov:1991rh} 
and the concept of  ${\rm b}$-quark invariant mass 
are playing an important role. In view of the fact
that the positive features of the invariant mass are sometimes not
taken used to the total extent, let us remind the basic steps of
definitions of this QCD parameter: 
\begin{enumerate} 
\item Define
the running quark mass through the solution of the following RG
equation for the anomalous dimension term (AD): 
\begin{equation}
\label{running} {\rm \overline{m}_b^2(\text{M}_\text{H})}
 ={\rm\overline{m}_b^2(m_b)}
  \exp\bigg[-2\int_{a_s(\text{m}_\text{b})}^{a_s(\text{M}_\text{H})}
\frac{\gamma_m(x)}{\beta(x)}dx\bigg]\,; 
\end{equation} 
\item Take
the integral in the r.h.s. of Eq.\ (\ref{running}): 
\begin{equation}
{\rm \overline{m}_b^2(\text{M}_\text{H})}= {\rm \overline{m}_b^2(m_b)} \bigg
(\frac{a_s(\text{M}_\text{H})}{a_s(\text{m}_\text{b})}\bigg)^{2\gamma_0/\beta_0} 
\bigg(\frac{AD(a_s(\text{M}_\text{H}))}{AD(a_s(\text{m}_\text{b}))}\bigg)^2~~~; 
\end{equation} 
\item Define
\begin{eqnarray}
 \nonumber
  AD(a_s)
   \!&\!=\!&\!
    \bigg[1+P_1a_s
           +\bigg(P_1^2+P_2\bigg)\,
             \frac{a_s^2}{2}
           +\bigg(\frac{1}{2}P_1^3+\frac{3}{2}P_1P_2+P_3\bigg)\,
             \frac{a_s^3}{3}
  \\ \label{AD}
   \!&\!{}\!&\!
           +\bigg(\frac{1}{6}P_1^4+\frac{4}{3}P_1P_3+P_1^2P_2+P_4\bigg)\,
             \frac{a_s^4}{4}
   \bigg]\,;
\end{eqnarray} 
\item Calculate its coefficients, 
 expressed through
\begin{equation*}
P_1=-\frac{\beta_1\gamma_0}{\beta_0^2}+\frac{\gamma_1}{\beta_0}
\approx 1.17549\,,~
P_2=\frac{\gamma_0}{\beta_0^2}\bigg(\frac{\beta_1^2}{\beta_0}-\beta_2\bigg)
-\frac{\beta_1\gamma_1}{\beta_0^2}+\frac{\gamma_2}{\beta_0} \approx
1.16196\,;~~~
\end{equation*} 
\begin{eqnarray}
 \nonumber
  P_3
   \!&\!=\!&\!
    \bigg[\frac{\beta_1\beta_2}
               {\beta_0}
         -\frac{\beta_1}{\beta_0}
           \bigg(\frac{\beta_1^2}{\beta_0}-\beta_2\bigg)
         -\beta_3
    \bigg]\,
     \frac{\gamma_0}{\beta_0^2} \\
   \!&\!{+}\!&\!
    \frac{\gamma_1}{\beta_0^2}
     \bigg(\frac{\beta_1^2}{\beta_0}-\beta_2\bigg)
   -\frac{\beta_1\gamma_2}{\beta_0^2}
   +\frac{\gamma_3}{\beta_0}
   \approx-3.1505 \\ \nonumber
  P_4
   \!&\!=\!&\!
    \frac{\gamma_0}{\beta_0^4}
     \bigg[\frac{\beta_1^2}{\beta_0^2}\,
            \bigg(\frac{\beta_1^2}{\beta_0}-\beta_2\bigg)
          +\frac{\beta_2^2}{\beta_0}
          -\frac{2\beta_1}{\beta_0}\,
            \bigg(\frac{\beta_1\beta_2}{\beta_0}-\beta_3\bigg)
          -\beta_4
     \bigg] \\ \nonumber
   \!&\!{+}\!&\!
    \frac{\gamma_1}{\beta_0^2}
     \bigg[\frac{\beta_1\beta_2}{\beta_0}
          -\frac{\beta_1}{\beta_0}\,
            \bigg(\frac{\beta_1^2}{\beta_0}-\beta_2\bigg)
          -\beta_3
     \bigg] \\
   \!&\!{+}\!&\!
    \frac{\gamma_2}{\beta_0^2}
     \bigg(\frac{\beta_1^2}{\beta_0}-\beta_2\bigg)
   -\frac{\gamma_3\beta_1}{\beta_0^2}
   +\frac{\gamma_4}{\beta_0}
   \approx-33.2389\,;
\end{eqnarray} 
\item Define the invariant mass 
\begin{equation} 
 \hat{\text{m}}_\text{b}
 =\overline{\text{m}}_\text{b}(\text{m}_\text{b})
   \bigg[a_s(\text{m}_\text{b})^{\frac{\gamma_0}{\beta_0}}
   {AD}(a_s(\text{m}_\text{b}))
   \bigg]^{-1}\,. 
\end{equation} 
It should be stressed, that this definition seems to be simpler 
than one, 
introduced in Ref.~\cite{Becchi:1980vz},
namely
\begin{equation} {\rm \hat{m}_b}={\rm
\overline{m}_b(m_b)}\bigg[(2\beta_0 a_s(\text{m}_\text{b}))^
{\frac{\gamma_0}{\beta_0}}{  AD(a_s(\text{m}_\text{b}))}\bigg]^{-1}\,,
\end{equation}
which is  often used in the literature. 
\end{enumerate} 
In the large-$\beta_0$ approximation the expression for
$\Gamma_{\rm{H\bar{b}b}}$ can be expressed as
~\cite{Broadhurst:2000yc} 
\begin{equation}
 \Gamma_{\rm{H\bar{b}b}}
  =\Gamma_0^{\text{b}}
    \frac{\rm\hat{m}_b^2(\text{M}_\text{H})}
         {\rm m_b^2}(a_s(\text{M}_\text{H}))^{\nu_0}
    \bigg[A_0+\sum_{n\geq1}{\rm d_n} A_n(a_s(\text{M}_\text{H}))
    \bigg] 
\end{equation} 
\begin{equation}
A_n=\frac{1}{\beta_0\delta_n\pi}\big(1+\beta_0^2\pi^2
a_s^2\big)^{-\delta_n/2} (a_s)^{n-1}{\sin}
\big[\delta_n{\arctan}(\beta_0\pi a_s)\big]\,, 
\end{equation} 
where
$\delta_n=n+\nu_0-1$, $\nu_0=2\gamma_0/\beta_0$. 
Taking now into account 
that within large-$\beta_0$ approximation, 
one has the following LO-expression
\begin{equation}
 \label{LO} 
 a_s(\text{M}_\text{H})_\text{LO}
  = \frac{1}{\beta_0\ln(\text{M}_\text{H}^2/\Lambda^2)}~~
\end{equation} 
fixing $n=0$ and expanding $A_0$  to first order in
$a_s$ we are getting the function: 
\begin{equation} 
A_0= \frac{1}{b\,L_{\text{M}_\text{H}}^b}
     \frac{{\sin}(b~{\arctan}(\pi/L_{\text{M}_\text{H}})}
          {(1+\pi^2/L_{\text{M}_\text{H}}^2)^{b/2}}\,,
\end{equation} 
where $b=\nu_0-1$,
$L_{\text{M}_\text{H}}=\rm{ln}(\text{M}_\text{H}^2/\Lambda^2)$. 
This expression was first derived in 80s 
in Ref.~\cite{Gorishnii:1983cu}. 
Unfortunately, its usefulness  was not understood at this time. 
At the new stage of the development of QCD the analogs of this formula, 
are forming the basis of FAPT method \cite{Bakulev:2006ex}, 
which is allowing to resum not only the terms 
proportional to $\gamma_0$ and $\beta_0$,
but higher order corrections of RG functions as well. 
Thus, the
extension of the Shirkov--Solovtsov method to the case of fractional
powers   \cite{Bakulev:2006ex} may  be considered in part as the
generalization for CIT resummation of  ``large-$\beta_0$''
contributions, which also  arises within  renormalon approach
~\cite{Broadhurst:2000yc}. In view  of the appearance of the works
of Refs.~\cite{Bakulev:2006ex,Stefanis:2008dz},~
\cite{Bakulev:2008qq}, it may be interesting to compare in the future
the results of these two approaches in more detail.

\section{On-shell and RG-resummation approach}
 \subsection{On-shell parameterization}
In the previous section we derived the relation between running and
invariant ${\rm b}$-quark masses.  However, there is also the
possibility to use  on-shell approach and express the width
$\Gamma_{\rm{H\bar{b}b}}$ through the ${\rm b}$-quark pole mass
${\rm m_b}$ and the ${\rm \overline{\text{MS}}}$-scheme coupling constant
$\alpha_s(\text{M}_\text{H})$ in different orders of perturbation theory and
compare the results obtained with the running mass motivated
RG-resummation approach. This analysis was done at the
$\alpha_s^2$-level in Refs.
\cite{Kataev:1992fe,Kataev:1993be}. 
In these studies the effects of mass dependent 
$O(\alpha_s{\rm m_b^2}/{\rm \text{M}_\text{H}^2})$-corrections, 
extracted from the calculations of Ref.~\cite{Surguladze:1990sp}, 
were also included. 
Among most important results of Refs.\cite{Kataev:1992fe,Kataev:1993be} 
were the explicit demonstration 
of the importance of taking into account
$\alpha_s^2$-corrections in the on-shell approach. 
Indeed, these effects turned out to be rather important 
for decreasing 
the difference between the $\Gamma_{\rm{H\bar{b}b}}$ expressions,
evaluated in the on-shell and RG-resummed approaches. 
The results of our studies were confirmed later on 
by the considerations of Ref.~\cite{Surguladze:1994gc}, 
where the expression 
of the $O(\alpha_s^2 {\rm m_b^2}/{\rm \text{M}_\text{H}^2})$-contribution
was evaluated and included. 
Note, however, that for  the considered at present masses 
of Higgs boson   
these effects are less important, 
than higher order perturbative QCD corrections, 
and can be safely neglected.
Keeping in mind the demands of Tevatron and LHC experiments
and the ongoing  discussion of the scientific program 
for International Linear Collider, 
in this section we will study the similar problem in more detail, 
taking into account the information  on available 
at present higher-order QCD corrections 
to the RG-functions, 
coefficient function for $\Gamma_{\rm{H\bar{b}b}}$ in Eq.\ (\ref{MS}) 
(see Sec.\,2) 
and the relation between running and on-shell masses
which we present in the following form: 
\begin{equation}
 \label{mb} 
 \frac{\rm\overline{m}_b^2(\rm{m_b})}
      {\rm m_b^2}
  = 1-\frac{8}{3}\,a_s(\text{m}_\text{b})
     -18.556\,a_s(\text{m}_\text{b})^2
     -175.76\,a_s(\text{m}_\text{b})^3
     -1892\,a_s(\text{m}_\text{b})^4\,.
\end{equation}
The $a_s^2$-correction is the
result of calculations of Ref.~\cite{Gray:1990yh}, 
confirmed later on in~\cite{Fleischer:1998dw}. 
The $a_s^3$-term  comes from the analytical  calculations 
of Ref.~\cite{Melnikov:2000qh} 
and is confirming semi-analytical similar result, 
obtained in Ref.~\cite{Chetyrkin:1999ys}. 
Note, that its  coefficient   
turned out to be in a good agreement 
with the  ECH/PMS estimate of Ref.~\cite{Chetyrkin:1997wm}. 
This fact and the success of  the ECH/PMS prediction 
for  the value of $\text{d}_4^\text{E}$ term
\cite{Baikov:2005rw}, we are supplementing    the estimates of
Ref.~\cite{Chetyrkin:1997wm} by
 definite RG-inspired considerations and get our personal
ECH-inspired number  for the coefficient of $a_s^4$-term in
Eq.\ (\ref{mb}). Proceeding further on with the help of the derived in
Ref.~\cite{Chetyrkin:1997wm} RG equations for the transformation of
$\rm{m_b^2(\text{M}_\text{H})}$ to  
$\overline{\text{m}}_\text{b}^2(\text{m}_\text{b})$ and of
$a_s(\text {m}_\text {b})$ to $a_s(\text{M}_\text{H})$, 
we get the following analog 
of Eq.\ (\ref{MS}):
\begin{equation} 
\label{OS}
 \Gamma_{\rm{H\bar{b}b}}
 =\Gamma_0^{\text{b}}
   \bigg[1
    + \Delta{\rm \Gamma^{b}_1} a_s(\text{M}_\text{H}) 
    + \Delta{\rm \Gamma^{b}_2} a_s(\text{M}_\text{H})^2
    + \Delta{\rm \Gamma^{b}_3} a_s(\text{M}_\text{H})^3 
    + \Delta{\rm \Gamma^{b}_3} a_s(\text{M}_\text{H})^4 
    \bigg],
\end{equation} 
where $\Gamma_0=3\sqrt{2}/(8\pi)\,G_F\,\text{M}_\text{H}\,\text{m}_\text{b}^2$ and 
\begin{eqnarray}
 \label{G1}
  \Delta{\rm \Gamma_1^{b}}
   \!&\!=\!&\! 3-2\,L\,; \\
 \label{G2}
  \Delta{\Gamma_2^{b}}
   \!&\!=\!&\!-4.5202
              -18.139\,L
              +0.08333\,L^2\,; \\
 \label{G3}
  \Delta{\Gamma_3^{b}}
   \!&\!=\!&\!-316.878
              -133.421\,L
              -1.15509\,L^2
              +0.050926\,L^3\,; \\
 \label{G4}
  \Delta{\Gamma_4^{b}}
   \!&\!=\!&\!-4366.17
              -1094.62\,L
              -55.867\,L^2
              -1.8065\,L^3
              +0.04774\,L^4~~~
\end{eqnarray} 
and  $L=\ln(\text{M}_\text{H}^2/\text{m}_\text{b}^2)$. We will
define now the   QCD coupling constant in different orders of
perturbation theory as 
\begin{eqnarray}
 \label{NLO}
  a_s(\text{M}_\text{H})_{\text{NLO}}
  \!\!&\!\!=\!\!&\!\!
   \frac{1}{\beta_0{\rm Log)}}
    \bigg[1-\frac{\beta_1{\ln(\text{Log})}}{\beta_0^2{\rm Log^2}}\bigg]\,; \\
 \label{N^2LO}
  a_s(\text{M}_\text{H})_{\text{N}^2\text{LO}}
   \!\!&\!\!=\!\!&\!\!
    a_s(\text{M}_\text{H})_{\text{NLO}}
    +\Delta a_s(\text{M}_\text{H})_{\text{N}^2\text{LO}}\,; \\
 \label{N^3LO}
  a_s(\text{M}_\text{H})_{\text{N}^3\text{LO}}
   \!\!&\!\!=\!\!&\!\!
    a_s(\text{M}_\text{H})_{\text{N}^2\text{LO}}
    +\Delta a_s(\text{M}_\text{H})_{\text{N}^3\text{LO}}\,;\\
 \label{N^4LO}
  a_s(\text{M}_\text{H})_{\text{N}^4\text{LO}}
   \!\!&\!\!=\!\!&\!\!
    a_s(\text{M}_\text{H})_{\text{N}^3\text{LO}}
    +\Delta a_s(\text{M}_\text{H})_{\text{N}^4\text{LO}}\,; \\
 \nonumber
  \Delta a_s(\text{M}_\text{H})_{\text{N}^2\text{LO}}
   \!\!&\!\!=\!\!&\!\!
    \frac{1}{\beta_0^5 \text{Log}^3}
     \bigg(\beta_1^2{\ln^2(\text{Log})}
          -\beta_1^2{\ln(\text{Log})}
          +\beta_2\beta_0
          -\beta_1^2
     \bigg)\,;\\ \nonumber
  \Delta a_s(\text{M}_\text{H})_{\text{N}^3\text{LO}}
   \!\!&\!\!=\!\!&\!\!
    \frac{1}{\beta_0^7\text{Log}^4}
     \bigg[\beta_1^3
            \bigg(-\ln^3(\text{Log})
                  +\frac{5}{2}{\ln^2(\text{Log})}
                  +2{\ln(\text{Log})}
                  -\frac{1}{2}
            \bigg) \\ \nonumber
   \!\!&\!\!\!\!&\!\!
    -3\beta_0\beta_1\beta_2{\ln(\text{Log})}
    +\beta_0^2\frac{\beta_3}{2}\bigg]\,; \\ \nonumber
 \Delta a_s(\text{M}_\text{H})_{\text{N}^4\text{LO}}
  \!\!&\!\!=\!\!&\!\!
   \frac{1}{\beta_0^9\text{Log}^5}
    \bigg[\beta_1^4
           \bigg({\ln^4(\text{Log})}
                -\frac{13}{3}{\ln^3(\text{Log})}
                -\frac{3}{2}{\ln^2(\text{Log})} 
                +4{\ln(\text{Log})} \\ \nonumber
  \!\!&\!\!\!\!&\!\!~~~~~~~~~~~~+\frac{7}{6}
                   \bigg)
         +3\beta_1^2\beta_2\,
           \bigg(2{\ln^2(\text{Log})}-{\ln(\text{Log})}-1\bigg) \\ \nonumber
  \!\!&\!\!\!\!&\!\!~~~~~~~~~~~~
         -\beta_1\beta_3\,
           \bigg(2{\ln(\text{Log})}
         +\frac{1}{6}\bigg)
         +\frac{5}{3}\beta_2^2
         +\frac{\beta_4}{3}
   \bigg]\,,
\end{eqnarray} 
where
$\text{Log}=2\,\ln(\text{M}_\text{H}/\Lambda^{({\rm n_f=5})}_{\overline{\text{MS}}})$ 
and the additional terms 
$\Delta a_s(\text{M}_\text{H})_{\rm N^3LO}$ 
and 
$\Delta a_s(\text{M}_\text{H})_{\rm N^4LO}$ 
were obtained in Refs.~\cite{Chetyrkin:1997sg,Kniehl:2006bg} 
with the corresponding $\rm{N^3LO}$ and
$\rm{N^4LO}$ matching conditions, which are allowing to determine
the values of $\Lambda^{({\rm n_f=5})}_{\overline{\text{MS}}}$ from
$\Lambda^{({\rm n_f=4})}_{\overline{\text{MS}}}$ by passing threshold of
production of heavy flavor (in our case on-shell mass $\rm{m_b}$).
The analytical results of Ref.~\cite{Kniehl:2006bg} are in 
complete agreement with the mixture of  previous analogous  analytical 
and semi-analytical calculations 
of Refs. \cite{Schroder:2005hy,Chetyrkin:2005ia}. 
These conditions generalize to higher orders  the $\rm{NLO}$ and
$\rm{N^2LO}$ formulae, derived in Ref.~\cite{Bernreuther:1981sg}
(the corresponding $\rm{N^2LO}$ relation was corrected in
Ref.~\cite{Larin:1994va}). To save the space, we will not present
here the explicit form of these equations. An interested reader can
consult Ref.~\cite{Kataev:2001kk}, where the results of Refs.
\cite{Bernreuther:1981sg,Larin:1994va} and
\cite{Chetyrkin:1997sg} are presented for the case of considering
${\rm b}$-quark pole mass as the matching point. The corresponding
${\rm N^4LO}$ version will be presented elsewhere.
In order to perform the on-shell analysis we are using the results
for the  ${\rm b}$-quark pole  mass values from
Ref.~\cite{Penin:2002zv}, which are increasing from LO to
$\rm{N^4LO}$, and the LO, $\rm{NLO}$, $\rm{N^2LO}$
 expressions for $\Lambda^{(\rm n_f=4)}_{\overline{\rm MS}}$, related to one of the most
recent parton distribution fits Ref.~\cite{Martin:2007bv}. It can be
shown, that NLO and $\rm{N^2LO}$ values of these parameters are in
agreement with the central values of 
$\Lambda^{(\rm n_f=4)}_{\overline{\rm MS}}$
obtained in the process of the performed in Ref.
\cite{Kataev:2001kk} fits of Tevatron $\nu N$ deep-inelastic
scattering data. In view of this the $\rm{N^3LO}$ value of
$\Lambda^{(\rm n_f=4)}_{\overline{\rm MS}}$  will be taken from Ref.
\cite{Kataev:2001kk}. Leaving aside the discussions of the possible
convergence of the fits, at the $\rm{N^4LO}$ we will use  the same
value of
  $\Lambda^{(\rm n_f=4)}_{\overline{\rm MS}}$ as at the $\rm{N^3LO}$ and will make the guess
about the value of the ${\rm b}$-quark mass. Using now the matching
conditions of Ref.~\cite{Kniehl:2006bg} transformed into the form,
given in Ref.~\cite{Kataev:2001kk}, we obtain the following values of
$\Lambda_{\overline{\text{MS}}}^{(\rm n_f=5)}$: 
\begin{table}[h]
\centerline{ \begin{tabular}{|c|p{20mm}|p{20mm}|p{20mm}|} \hline
order  & ${\rm m_b}~\text{GeV}$ & $\Lambda^{(\rm
n_f=4)}_{\overline{\rm MS}}$~${\rm MeV}$ &
 $\Lambda^{(\rm n_f=5)}_{\overline{\rm MS}}$~${\rm MeV}$ \\
\hline
LO & 4.74 & 220 & 168 \\
NLO & 4.86 & 347 & 254 \\
${\rm N^2LO}$  & 5.02     & 331    & 242 \\
${\rm N^3LO}$  & 5.23     & 333    & 243 \\
${\rm N^4LO}$  & 5.45     & 333    & 241 \\
\hline \end{tabular}} \caption{The values of the parameters on the
on-shell studies.}
\label{Tab} 
\end{table} 

In order to study the
effects of separate NLO, $\rm{N^2LO}$, $\rm{N^3LO}$ and $\rm{N^4LO}$
corrections to different parameterizations,  it is also useful to
define the quantity 
\begin{equation} 
 R_{\rm{H\bar{b}b}}
 = \Gamma_{\rm{H\bar{b}b}}/\Gamma_0^{\text{b}}
\end{equation} 
and study its $\text{M}_\text{H}$ dependence. At the final
step it is necessary  to substitute the results from Table 1 into
Eq.\ (\ref{OS})--(\ref{N^4LO}) by iterative way and compare
necessary curves. 
They are shown in Fig.\ 1 
both for $\Gamma_{\rm{H\bar{b}b}}$ 
and $R_{\rm {H\bar{b}b}}$ quantities.

\subsection{Renormalization group resummation approach}
In order to apply RG-resummation approach to
$\Gamma_{\rm{H\bar{b}b}}$ 
it is necessary to combine definition of Eq.\ (1), 
the basic formulae (27)--(29), 
the concrete numbers of Eqs.\ (30)--(31), 
the transformation relation of Eq.\ (38) and the
expressions for $a_s(\text{M}_\text{H})$ 
defined through Eqs.\ (44)--(47) 
and the similar expressions for $a_s(\text{m}_\text{b})$. 
The results are presented at Fig.\ 2. 
\newpage
\begin{figure}[h] 
{\includegraphics[width=65mm]{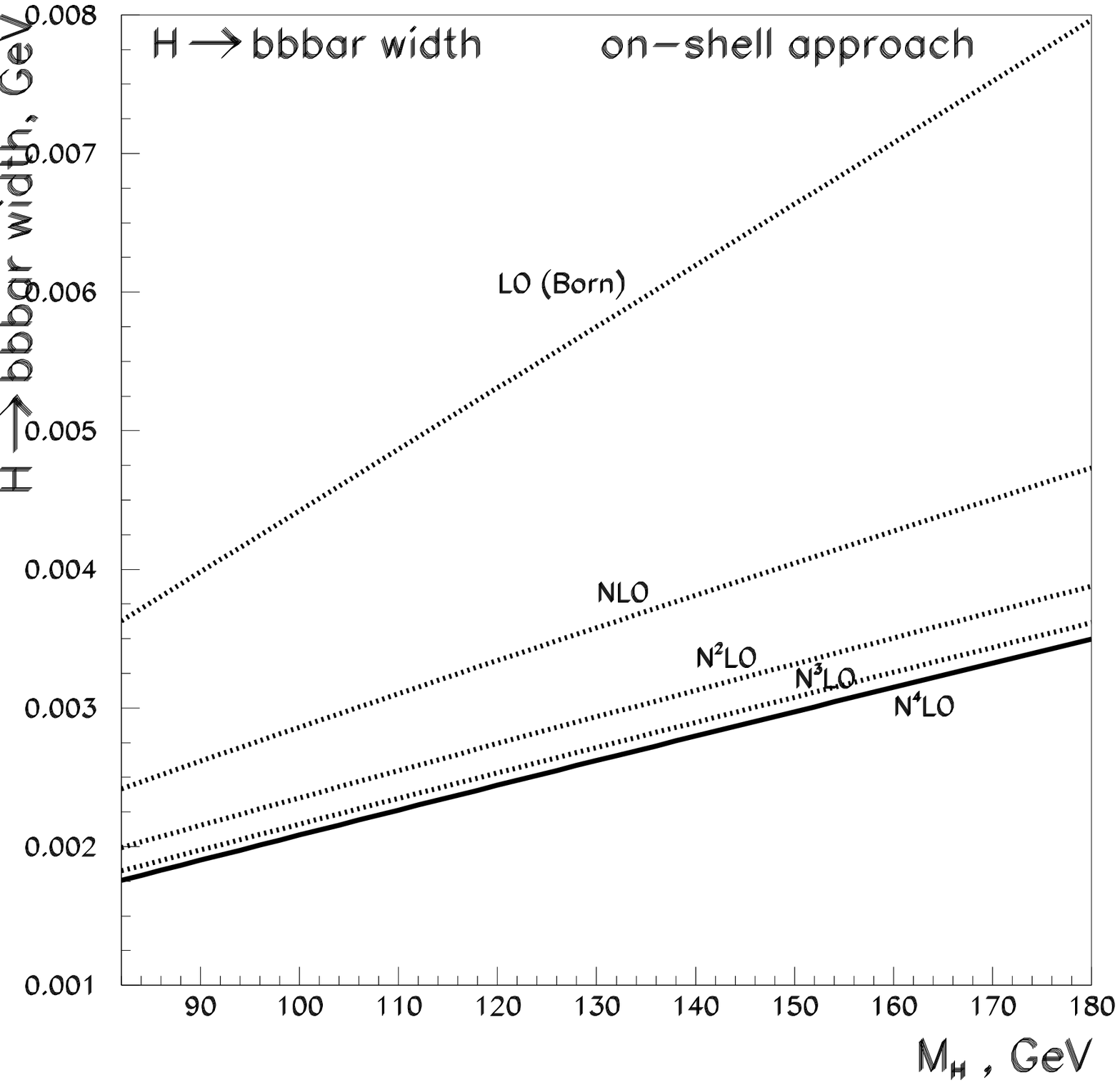}
 \includegraphics[width=65mm]{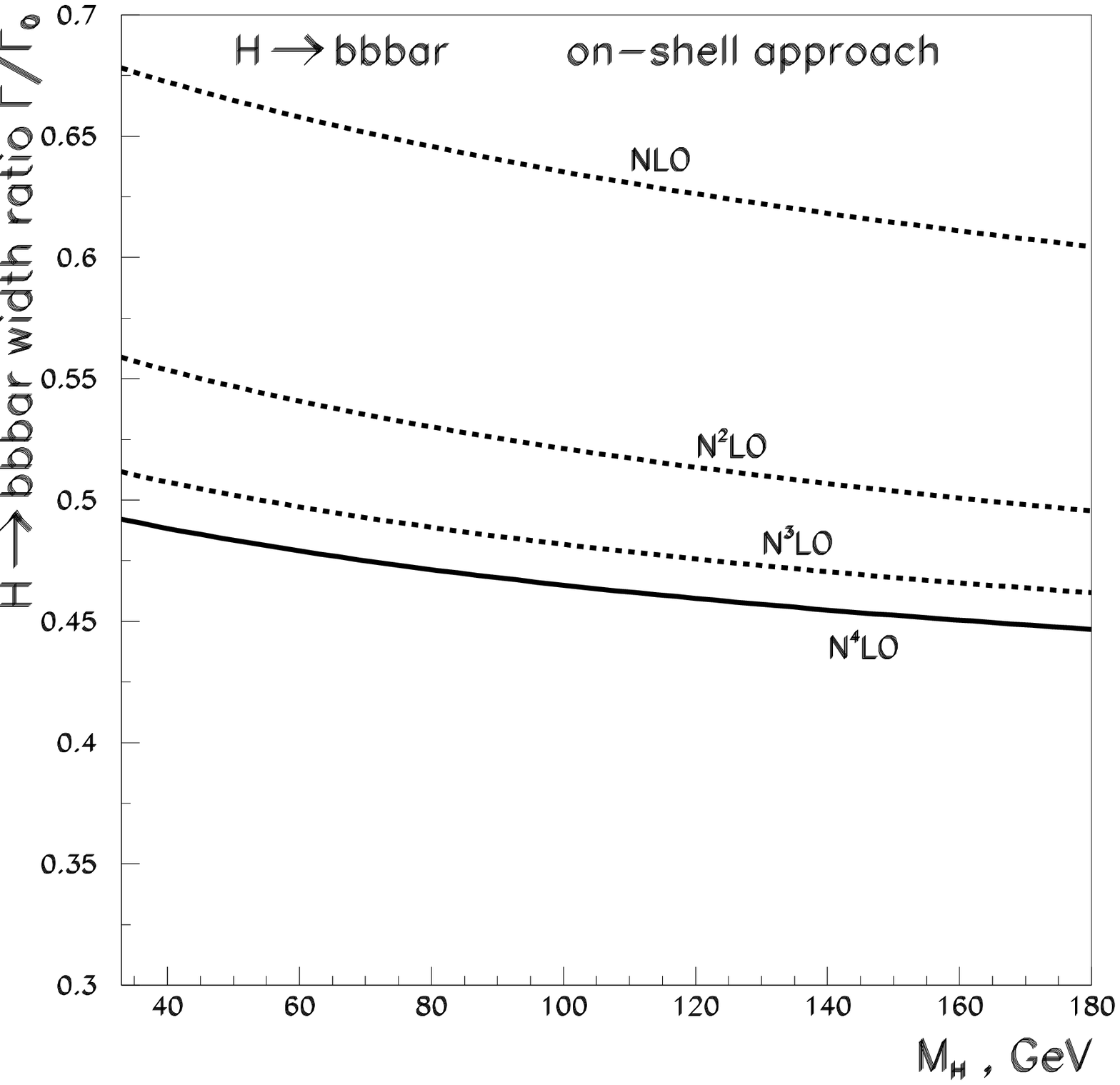}}
 \caption{The analyzed quantities  in the OS-approach. 
 \label{fig1}}
{\includegraphics[width=65mm]{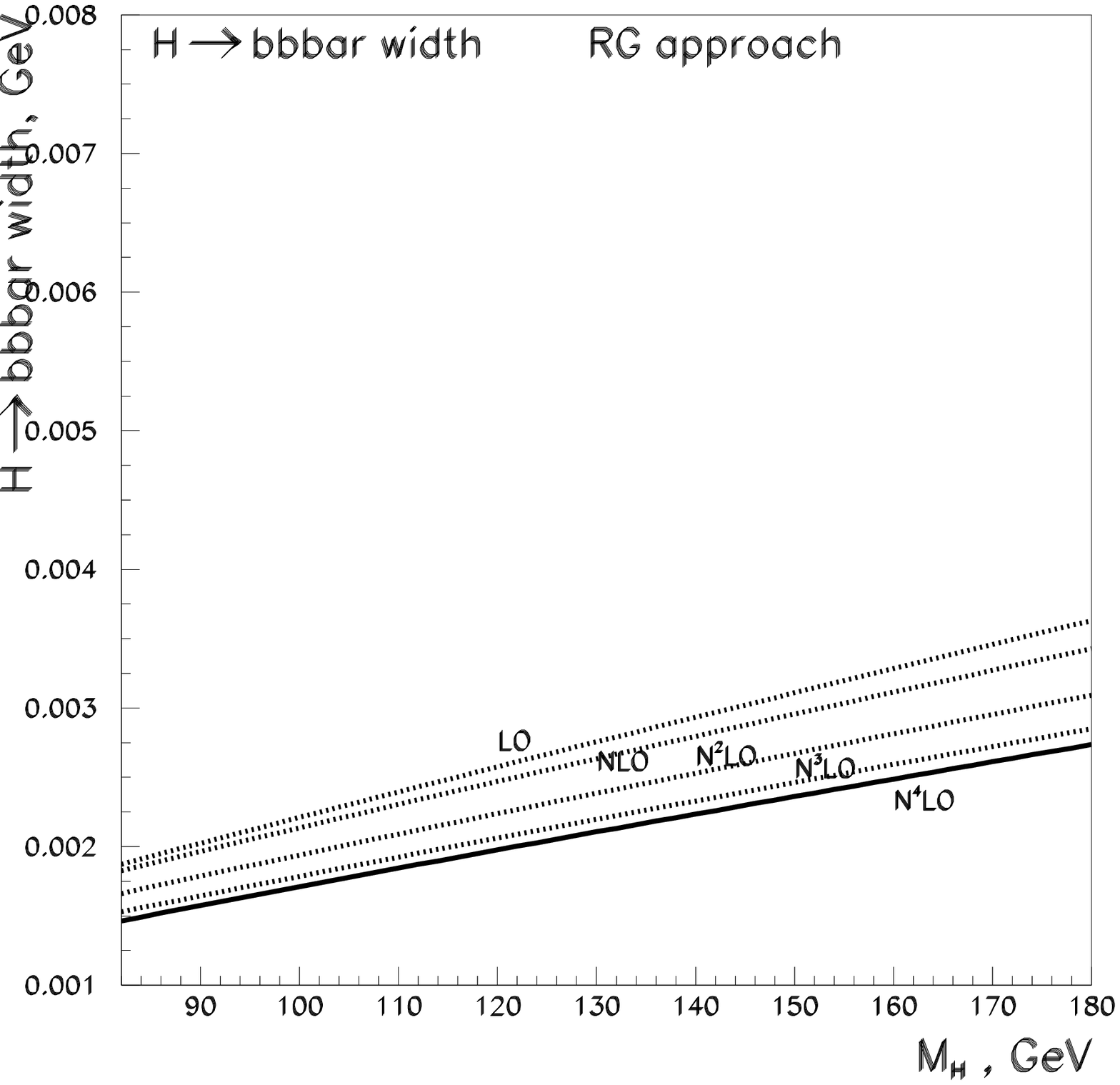}
 \includegraphics[width=65mm]{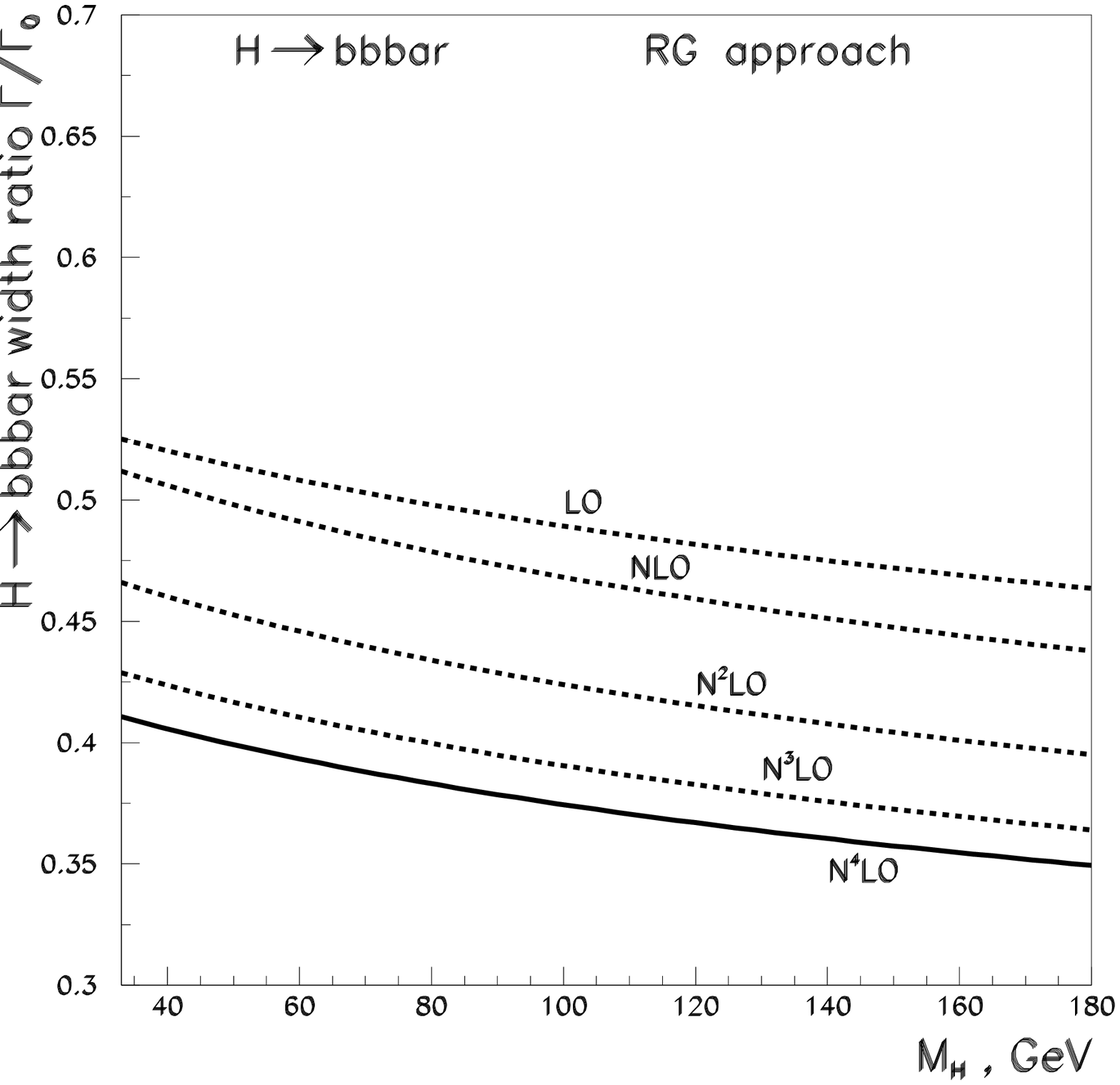}}
  \caption{The analyzed quantities in the RG-approach. 
  \label{fig2}} 
\end{figure}
$\Gamma_{\rm {H\bar{b}b}}$ may be  presented in the following form
\begin{eqnarray} \label{RG}
 \Gamma_{\rm {H\bar{b}b}}
 \!&\!=\!&\!
 \Gamma_0^{(b)}
  \bigg(\frac{a_s(\text{M}_\text{H})}
             {a_s(\text{m}_\text{b})}
  \bigg)^{(24/23)}
   \frac{AD(a_s(\text{M}_\text{H}))^2}
        {AD(a_s(\text{m}_\text{b}))^2} \bigg[1+\sum_{i\geq 1}
              \Delta{\rm \Gamma_i}\,
               a_s^i(\text{M}_\text{H})
     \bigg]
  \\ \nonumber 
  \!&\!\times\!&\!
  \big(1-\frac{8}{3}\,a_s(\text{m}_\text{b})
        -18.556\,a_s(\text{m}_\text{b})^2
        -175.76\,a_s(\text{m}_\text{b})^3
        -1892.2\,a_s(\text{m}_\text{b})^4
  \big)\,,
\end{eqnarray}
where
\begin{equation}
 AD(a_s)^2=1+2.351\,a_s^2+4.383\,a_s^2+3.873\,a_s^3-15.153\,a_s^4
\end{equation}
Different curves at Figures 1 and 2 are related to applications of
the results for $\Gamma_{\rm {H\bar{b}b}}$ and $R_{\rm
{H\bar{b}b}}$ of the step-by-step substitution into the
definitions of the corresponding expressions for the QCD
coupling constants $a_s(\text{M}_\text{H})$ and $a_s(\text{m}_\text{b})$ 
from Eqs.\ (36),
(44)--(47).
The considerations of  Fig.~1 confirm the findings of Refs.
\cite{Kataev:1992fe,Kataev:1993be} on the importance of the
1- and 2-loop contributions to
 $\Gamma_{\rm {H\bar{b}b}}$ in parameterization with on-shell mass.
On the other hand, the other two  corrections are not so big, but they
both have a tendency to decrease the value of $\Gamma_{\rm
{H\bar{b}b}}$ to its RG-improved expression (see Fig.~2).
Moreover, the RG-resummation approach demonstrate clearly, that the
perturbative theory to this quantity are well under control. The
r.h.s. parts of Fig.~1 and Fig.~2 demonstrate the stability of good
convergence of the related perturbative approximations. Thus, the
consideration of the order $\alpha_s^4$ corrections, calculated in
Ref.~\cite{Baikov:2005rw}, support the feature of minimizing the
difference between on-shell and RG-resummed parameterizations,
already observed at the $\alpha_s^3$-level in Ref.
\cite{Kataev:2007kf}. The l.h.s. plots of Fig.~1 and Fig.~2 are more
interesting from phenomenological point of view. It will be the next
task to compare these results for the Higgs decay width, obtained by
using FAPT approach of Ref.~\cite{Bakulev:2006ex} and resummed FAPT
analysis of Ref.~\cite{Bakulev:2008qq}. Thus, our considerations
leave space for further studies of peculiar features 
(presumably,  better
convergence) of modified APT predictions.

\section{Acknowledgements}
This work is dedicated to the memory of our friend and colleague
I.~L.~Solovtsov (09.01.1952--28.07.2007). He left this world too
early. However, he managed to create  the ideas and important  works,
which are  definitely useful for us at present 
and let us hope will be still inspiring in the future.
The work of one of us (A.L.K) was done within the framework of RFBR
Grants 08-01-00686-a and 06-02-16659-a.  The work of V.T.K is
supported in parts by  RFBR Grants 08-02-01184-a and
06-02-72041-MNTI-a. We are grateful to A.~P.~Bakulev for discussions.


\end{document}